\documentclass[pra,aps,twocolumn]{revtex4-1}
\usepackage{graphicx}
\usepackage{amsmath}
\usepackage{bm}

\newcommand{\be}{\begin{equation}}
\newcommand{\ee}{\end{equation}}
\newcommand{\bea}{\begin{eqnarray}}
\newcommand{\eea}{\end{eqnarray}}
\newcommand{\bei}{\begin{itemize}}
\newcommand{\eei}{\end{itemize}}

\let\oldhat\hat
\renewcommand{\vec}[1]{\mathbf{#1}}
\renewcommand{\hat}[1]{\oldhat{#1}}

\begin{document}
\title{Zero sound in a two-dimensional dipolar Fermi gas}

\author{ Zhen-Kai Lu$^{1}$, S.I. Matveenko$^{2,3}$, and G. V. Shlyapnikov$^{2,4,5}$}
\affiliation{
\mbox{$^{1}$Max-Planck-Institut f\"ur Quantenoptik, Hans-Kopfermann-Stra{\ss}e 1, 85748 Garching, Germany}\\ 
\mbox{$^{2}$Laboratoire de Physique Th\'eorique et Mod\`eles Statistiques, CNRS and Universit\'e Paris Sud, UMR8626, 91405 Orsay, France}\\
\mbox{$^{3}$L.D. Landau Institute for Theoretical Physics, Kosygina Str. 2, 119334, Moscow, Russia}\\ 
\mbox{$^{4}$Van der Waals-Zeeman Institute, University of Amsterdam, Science Park 904, 1098 XH Amsterdam, The Netherlands}\\
\mbox{$^{5}$ Kavli Institute for Theoretical Physics, University of California, Santa Barbara, CA 93106-4030, USA}}
\begin{abstract}

We study zero sound in a weakly interacting 2D gas of single-component fermionic dipoles (polar molecules or atoms with a large magnetic moment) tilted with respect to the plane of their translational motion. It is shown that the propagation of zero sound is provided by both mean field and many-body (beyond mean field) effects, and the anisotropy of the sound velocity is the same as the one of the Fermi velocity. The damping of zero sound modes can be much slower than that of quasiparticle excitations of the same energy. One thus has wide possibilities for the observation of zero sound modes in experiments with 2D fermionic dipoles, although the zero sound peak in the structure function is very close to the particle-hole continuum.  

\end{abstract}
\date{\today}
\pacs{}
\maketitle

\section{Introduction}

The creation of quantum gases of atoms with large magnetic moments \cite{Pfau,Lahaye2009,Lev,Ferlaino} and ultracold clouds of ground state diatomic polar molecules \cite{Ni,Carr} strongly stimulated the work in the domain of dipolar cold gases, because the long-range anisotropic dipole-dipole interaction drastically changes the nature of quantum degenerate regimes. Presently, there is a growing number of proposals to study new classes of many-body states in these systems \cite{Baranov2008,Pupillo2008,Baranov2012}. A serious difficulty for studying many-body physics with polar molecules is related to ultracold chemical reactions, such as KRb+KRb$\Rightarrow$K$_2$+Rb$_2$ observed in the JILA experiments \cite{Jin,Jin2}, which lead to a rapid decay of the system at required densities. Therefore, the attention is now shifted to non-reactive molecules, in particular RbCs, NaK, and KCs, for which the ultracold chemistry is expected to be energetically unfavorable \cite{Jeremy}. Another route assumes the suppression of chemical reactions for reactive molecules by confining them to the (quasi)two-dimensional (2D) geometry and orienting their dipole moments (by a strong electric field) perpendicularly to the plane of the 2D translational motion \cite{Bohn1, Baranov1}. This induces a strong intermolecular repulsion, and the suppression of chemical reactions by two orders of magnitude has been already demonstrated \cite{Ye}.

Therefore, 2D gases of polar molecules attract a special attention, in particular when the molecules are fermionic and they are in the same internal state. One then has an additional reduction of chemical reactions. Various aspects have been discussed regarding this system in literature, in particular the emergence and beyond mean field description of the topological $p_x+ip_y$ phase for microwave-dressed polar molecules \cite{Cooper2009,Gora}, interlayer superfluids in bilayer and multilayer systems \cite{Pikovski2010,Ronen,Potter,Zinner}, the emergence of density-wave phases \cite{Sun,Miyakawa,Parish,Babadi,Baranov}, and superfluid pairing for tilted dipoles  \cite{Taylor,Baranov}. The Fermi liquid behavior of this system has been addressed by using the Fourier transform of the dipole-dipole interaction potential (see \cite{Baranov} and refs. therein). The many-body theory (beyond mean field) describing Fermi liquid properties of a weakly interacting 2D gas of identical fermionic dipoles with dipole moments $d$ oriented perpendicularly to the plane of their translational motion, has been developed in Ref.~\cite{Lu}. The theory relies on the presence of a small parameter $p_Fr_*$, where $p_F$ is the Fermi momentum, and $r_*=md^2/\hbar^2$ is the dipole-dipole length, with $m$ being the molecule mass. With the use of the low-momentum solution of the scattering problem up to terms $\sim (pr_*)^2$, thermodynamic quantities were obtained as a series of expansion up to the second order in $p_Fr_*$. Recent Monte Carlo calculations \cite{Giorgini} confirmed the findings of Ref.~\cite{Lu} in the low-density limit ($p_Fr_*<1$) and studied a quantum transition to the crystalline phase at high densities.     

In some sense, the 2D gas of identical fermionic dipoles perpendicular to the plane of their translational motion, constitutes a novel Fermi liquid because the existence of zero sound in this system is provided only by many-body effects \cite{Lu}. This stimulates an interest to zero sound for tilted dipoles and to possibilities for the observation of zero sound modes in experiments. In the present paper we show that in the 2D gas of identical fermionic tilted dipoles the propagation of zero sound is due to both mean field and many-body effects. The sound modes are anisotropic, and the anisotropy of the sound velocity is the same as the one of the Fermi velocity. Importantly, the damping rate of zero sound can be much lower compared to the damping rate of quasiparticle excitations with the same energy. This is different from the situation in $^3$He \cite{Abrikosov1958}, where these damping rates are of the same order of magnitude. The small damping rate of zero sound in the 2D dipolar Fermi gas opens wide possibilities for the observation of the sound modes in experiments, in spite of the fact that the zero sound peak is very close to the particle-hole continuum in the structure function.    

The paper is organized as follows. In Section II we give general relations for various quantities of the 2D gas of tilted fermionic dipoles and in Section III we derive the many-body contribution to the interaction function of quasiparticles in this system. Section IV is dedicated to the derivation of our results for the dynamical structure factor and zero sound velocity. It is in particular shown that for tilted dipoles the second order mean field contribution to this quantity and the many-body contribution are of the same order of magnitude. In Sections V and VI we calculate the relaxation rate of quasiparticles and the damping rate of zero sound, showing that the latter can be much smaller at the same excitation energy. We conclude in Section VII, emphasizing that the slow damping of zero sound provides wide possibilities for the measurement of zero sound modes. Aside from the observation of the surface modes in trapped samples, which in the collisionless regime are analogous to zero sound and have been observed in the 2D atomic Fermi gas \cite{Kohl}, one should be able to observe zero sound in the response  to small modulations of the density in (quasi)uniform gases like those created in the recent experiment \cite{Hadzibabic}. In contrast to experiments in liquid $^3$He, where the observation of zero sound is based on the difference between the zero sound and Fermi velocities \cite{heliumexp}, in ultracold gases the zero sound can be observed through the measurement of the dynamical structure factor in two-photon Bragg spectroscopy experiments. This method was successfully developed for Bose-condensed gases \cite{Bragg1,Bragg2} and then used for ultracold fermions \cite{Bragg3}. Although the zero sound peak in the structure function is located very close to the particle-hole continuum, it can be visible as it may be higher than the maximum of the continuum due to slow damping of the zero sound.    

\section{General relations. Anisotropy of the Fermi surface}

We consider a 2D gas of single-component fermionic dipoles tilted by an angle $\theta_0$ with respect to the plane of their translational motion (see Fig.~1). These dipolar particles interact with each other via the potential which at large separations $r$ is 
\begin{equation}  \label{Ur}
U({\bf r})=\frac{d^2}{r^3}(1-3\sin^2\theta_0\cos^2\theta), 
\end{equation}
where $\theta$ is the angle between the vector ${\bf r}$ and the $x$ axis in which the dipoles are tilted. The Hamiltonian of the system reads:
\begin{equation}
\label{H}
\hat{\cal H}=\sum_{\vec{p}}\xi_p\hat{a}_{\vec{p}}^{\dag}\hat{a}_{\vec{p}}+\frac{1}{2S}\!\!\!\sum_{\vec{p_{1}},\vec{p_{2}},\vec{q}}\!\!\!U(\vec{q}) \hat{a}_{\vec{p_{1}}+\vec{q}}^{\dag}\hat{a}_{\vec{p_{2}}-\vec{q}}^{\dag}\hat{a}_{\vec{p_{2}}}\hat{a}_{\vec{p_{1}}},
\end{equation}
where $S$ is the surface area, $\xi_p=\hbar^2p^2/2m-\mu$ with $\mu$ being the chemical potential, $\hat {a}^{\dagger}_{{\bf p}}$ and $\hat{ a}_{{\bf p}}$ are creation and annihilation operators of fermionic dipoles with momentum ${\bf p}$, and $U(\vec{q})$ is the Fourier transform of the interaction potential $U({\bf r})$: 
\begin{equation}   \label{Uq}
U(\vec q)=\int d^2\vec{r} U({\bf r})e^{-i\vec{q}\cdot\vec{r}},
\end{equation} 
We focus on the weakly interacting regime, where the interaction energy per particle is much smaller than the Fermi energy and the inequality 
\begin{equation}   \label{kFr}
p_Fr_*\ll 1
\end{equation}
is satisfied. 

\begin{figure}
\includegraphics[width=9cm]{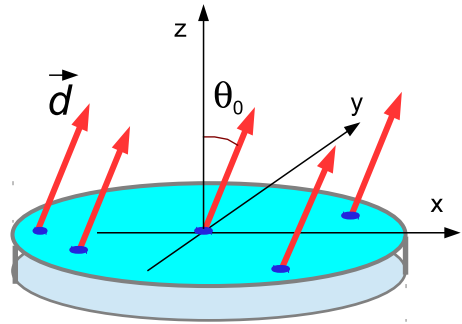}
\caption{(color online) 2D gas of dipoles tilted by an angle $\theta_0$ in the $x,z$-plane.}
\end{figure}

The potential $U({\bf r})$ becomes partially attractive at sufficiently large tilting angles $\theta_0$, providing a possibility of superfluid pairing. This occurs at $\theta_0$ exceeding a critical value $0.72$ \cite{Baranov}.  Assuming the absence of superfluid pairing the ground state of the system is a Fermi liquid and one may use the Landau theory relying on the existence of ''dressed particles'', or quasiparticles. At $T=0$ the momentum distribution of free quasiparticles is the step function
\begin{equation}   \label{nstep}
n({\bf p})=\theta(p_F-p),
\end{equation}
i.e. $n({\bf p})=1$ for $p<p_F$ and zero otherwise.The chemical potential is equal to the boundary energy at the Fermi circle, $\mu=\epsilon_F\equiv\epsilon(p_F)$. The quasiparticle energy $\epsilon({\bf p})$ is a variational derivative of the total energy with respect to the distribution function $n({\bf p})$. Due to the interaction between quasiparticles, the deviation $\delta n$ of this distribution from the step function (\ref{nstep}) results in a change of the quasiparticle energy:
\begin{equation}
\label{1}
\delta\epsilon (\vec{p})=\int F(\vec{p},\vec{p}')  \delta n(\vec{p}') \frac{d^2p'}{(2\pi)^2}.
\end{equation}
The interaction function of quasiparticles $F(\vec{p},\vec{p}')$ is thus the second variational derivative of the total energy with regard to $n({\bf p})$. The quantity $\delta n({\bf p})$ is significantly different from zero only near the Fermi surface, so that one may put ${\bf p}=p_F{\bf n}$ and ${\bf p}'=p_F{\bf n}'$ in the arguments of $F$ in Eq.~(\ref{1}), where ${\bf n}$ and ${\bf n}'$ are unit vectors in the directions of ${\bf p}$ and ${\bf p}'$. 

The knowledge of the interaction function of quasiparticles allows one to calculate the compressibility relying only on the integration on the Fermi surface. One then obtains straightforwardly the chemical potential, ground state energy, and other thermodynamic quantities. This idea belongs to Landau \cite{Landau56} and it was pushed forward by Abrikosov and Khalatnikov \cite{Abr} and implemented for a two-component 3D Fermi gas with a weak contact repulsion, revealing many-body (beyond mean field) effects and reproducing the results of an earlier (direct) calculation of Lee-Huang-Yang \cite{Lee,Huang}. The many-body theory relying on the Abrikosov-Khalatnikov approach has been developed in Ref.~\cite{Lu} for 2D fermionic single-component dipoles perpendicular to the plane of their translational motion ($\theta_0=0$). This theory accounts for the short-range physics in the scattering properties and represents thermodynamic quantities as a series of expansion in the small parameter $p_Fr_*$ up to the second order. The construction of a similar theory for tilted dipoles requires extremely cumbersome calculations and is beyond the scope of the present paper. Instead, we intend to reveal the properties of zero sound.

A distinguished feature of tilted 2D fermionic dipoles is a small anisotropy of the Fermi surface. The related corrections to the Fermi momentum, chemical potential, etc., are proportional to $p_Fr_*$, and we omit higher order corrections. Relations for these quantities have been derived in Ref.~\cite{Baranov}, and we present them here for completeness without going into detailed calculations. In contrast to strongly interacting systems, in the weakly interacting regime the quasiparticle energy is well defined at any momenta, not only near the Fermi surface. One may write:
\begin{equation}     \label{epsilonk}
\epsilon({\bf p})=\frac{\hbar^2p^2}{2m}-\mu+\int F({\bf p},{\bf p}')n({\bf p}')\frac{d^2p'}{(2\pi)^2},
\end{equation}  
and take into account that exactly on the Fermi surface the quasiparticle energy is zero:
\begin{equation}     \label{epsilonkF}
\epsilon({\bf p_F})=\frac{\hbar^2p_F^2}{2m}-\mu+\int F({\bf p_F},{\bf p}')n({\bf p}')\frac{d^2p'}{(2\pi)^2}=0.
\end{equation} 
In order to express the Fermi momentum through the density $n$ and the interaction strength, one has to solve equations (\ref{epsilonk}) and (\ref{epsilonkF}) selfconsistently with the particle number equation
\begin{equation}   \label{n}
n=\int\frac{d^2p}{(2\pi)^2}n({\bf p}).
\end{equation}
Integrating the right-hand side of this equation one has
\begin{equation}    \label{intphi}
\int\frac{p_F^2(\phi)d\phi}{2\pi}=p_{F0}^2=4\pi n,
\end{equation}
where $\phi$ is the angle between the vector ${\bf p}_F$ and the $x$ axis. Turning back to equation (\ref{epsilonkF}) we notice that to linear order in $p_Fr_*$ one may use the distribution function of a non-interacting Fermi gas $n(p')=\theta(p_{F0}-p')$ and put the interaction function of quasiparticles expressed through the Fourier transforms of the interaction potential $U({\bf r})$ \cite{Baranov,Lu}:
\begin{eqnarray}     
&&F_1\!({\bf p},\!{\bf p}')=[U(0)-U({\bf p}-{\bf p}')]  \nonumber \\
&&=2\pi d^2|{\bf p}-{\bf p}'|(\cos^{2}\theta_0-\sin^{2}\theta_0\cos^{2}\phi_{{\bf p}-{\bf p}'}), \label{F1} 
\end{eqnarray}
where $\phi_{{\bf p}-{\bf p}'}$ is the angle between the vector ${\bf p}-{\bf p}'$ and the $x$ axis. Then, integrating Eq.(\ref{epsilonkF}) over $d\phi$ and using Eq.(\ref{intphi}) we have:
\begin{equation}     \label{muanis}
\mu=\epsilon_{F0}\left(1+\frac{32}{9\pi}p_Fr_*P_2(\cos\theta_0)\right),
\end{equation}
with $\epsilon_{F0}=\hbar^2p_{F0}^2/2m$, and $P_2(\cos\theta_0)=(3\cos^2\theta_0-1)/2$ being the second order Legendre polynomial. Eq.(\ref{epsilonkF}) immediately gives the anisotropic Fermi momentum:
\begin{equation}      \label{kFanis}
p_F(\phi)=p_{F0}\left(1+\frac{8}{15\pi}p_Fr_*\sin^2\theta_0\cos 2\phi\right).
\end{equation}
For the quasiparticle energy near the Fermi surface, from Eq.(\ref{epsilonk}) we have:
\begin{equation}     \label{epsilonphi}
\epsilon({\bf p})=v_F(\phi)(p-p_F(\phi)),
\end{equation}
where 
\begin{equation}     \label{vFanis}
\!\!\!v_F(\!\phi\!)\!\!=\!\!v_{F0}\!\left(\!\!1\!\!+\!\!\frac{4}{3\pi}p_Fr_*P_2(\!\cos\theta_0\!)\!\!-\!\!\frac{2}{5\pi}p_Fr_*\sin^{\!2}\!\theta_0\cos 2\phi\!\!\right)\!\!\!\!\!
\end{equation}
is the radial component of the Fermi velocity, and $v_{F0}=\hbar p_{F0}/m$.

\section{Interaction function of quasiparticles}

The interaction function of quasiparticles $F({\bf p},{\bf p}')$ consists of two parts: the mean field part and the many-body one, and we need to know this function on the Fermi surface ($|{\bf p}|=|{\bf p}'|=p_F$). The mean field term is expressed through the scattering amplitude \cite{Lu}, and the contribution linear in $pr_*$ and $p'r_*$ is given by Eq.(\ref{F1}). The contribution $F_2({\bf p},{\bf p}')$ which is quadratic in $pr_*$ and $p'r_*$, depends on the short-range physics. It is obtained from the solution of the 2D scattering problem, and we omit terms which are proportional to higher powers of $pr_*$ and $p'r_*$ (see \cite{Lu} for $\theta_0=0$). Thus, the mean field contribution to the interaction function can be written as 
\begin{equation}   \label{Fmf}
F_{mf}({\bf p},{\bf p}')=F_1({\bf p},{\bf p}')+F_2({\bf p},{\bf p}'),
\end{equation}
with $F_1({\bf p},{\bf p}')$ from Eq.(\ref{F1}). We do not specify the expression for $F_2({\bf p},{\bf p}')$ and only mention that $F_2=0$ for ${\bf p}={\bf p}'$. 

The many-body part of the interaction function is obtained as the second variational derivative of the many-body (beyond mean field) contribution to the total energy. This contribution is expressed in terms of the off-shell scattering amplitude $f({\bf p}',{\bf p})$ \cite{Lu}:
\begin{eqnarray}   
&&\tilde E_{mb}=-\frac{1}{2S^2}\sum_{\vec{p_{1}},\vec{p_{2}},\vec{p'_{1}}}\frac{2m|f(\vec{p'},\vec{p})-f({\bf p'},-{\bf p})|^2}{\hbar^2(\vec{p^{2}_{1}}+\vec{p^{2}_{2}}-\vec{p'^{2}_{1}}-\vec{p'^{2}_{2}})}  \nonumber  \\
&&\times n({\bf p}_1)n({\bf p}_2)n({\bf p}'_1)\delta_{{\bf p}_1+{\bf p}_2-{\bf p}'_1-{\bf p}'_2}, \label{Emb}
\end{eqnarray}
where ${\bf p}=({\bf p}_1-{\bf p}_2)/2$ and ${\bf p}'=({\bf p}'_1-{\bf p}'_2)/2$. As we are interested only in the powers of $pr_*$ and $p'r_*$ not larger than two, the small anisotropy of the Fermi surface can be omitted. For the off-shell amplitudes in Eq.(\ref{Emb}) we may take the result of the first Born approximation which is linear in $pr_*$ and $p'r_*$:
\begin{widetext}
\begin{equation}     \label{ff}
f({\bf p}',{\bf p})-f({\bf p}',-{\bf p})=2\pi d^2\left\{|{\bf p}'+{\bf p}|(\cos^2\theta_0-\sin^2\theta_0\cos^2\phi_{{\bf p}+{\bf p}'})-|{\bf p}-{\bf p}'|(\cos^2\theta_0-\sin^2\theta_0\cos^2\phi_{{\bf p}-{\bf p}'})\right\}.
\end{equation}
We then obtain the many-body contribution to the interaction function:
\begin{eqnarray}
&&F_{mb}({\bf p},{\bf p}')=\tilde F_1({\bf p},{\bf p}')+\tilde F_2({\bf p},{\bf p}'); \label{tildeF}  \\
&&\tilde F_1({\bf p},{\bf p}')=-\frac{4\hbar^2}{m}(p_{F0}r_*)^2\int_{p_1<p_{F0}}\frac{d^2p_1}{p_{F0}^2}\frac{\delta_{{\bf p}+{\bf p}_1-{\bf p}'-{\bf p}_2}}{p_1^2-p_2^2}\Big\{\cos^2\theta_0(|{\bf p}'-{\bf p}|-|{\bf p}'-{\bf p}_1|) \nonumber \\ 
&&+\sin^2\theta_0\left[\frac{(p'\cos\phi_{{\bf p}'}-p_1\cos\phi_{{\bf p}_1})^2}{|{\bf p}'-{\bf p}_1|}-\frac{(p'\cos\phi_{{\bf p}'}-p\cos\phi_{{\bf p}})^2}{|{\bf p}-{\bf p}'|}\right]\Big\}^2; \label{tildeF1} \\
&&\tilde F_2({\bf p},{\bf p}')=-\frac{2\hbar^2}{m}(p_{F0}r_*)^2\int_{p_1<p_{F0}}\frac{d^2p_1}{p_{F0}^2}\frac{\delta_{{\bf p}+{\bf p}'-{\bf p}_1-{\bf p}_2}}{2p_{F0}^2-p_1^2-p_2^2}\Big\{\cos^2\theta_0(|{\bf p}_1-{\bf p}|-|{\bf p}'-{\bf p}_1|) \nonumber \\ 
&&+\sin^2\theta_0\left[\frac{(p'\cos\phi_{{\bf p}'}-p_1\cos\phi_{{\bf p}_1})^2}{|{\bf p}'-{\bf p}_1|}-\frac{(p_1\cos\phi_{{\bf p}_1}-p\cos\phi_{{\bf p}})^2}{|{\bf p}-{\bf p}_1|}\right]\Big\}^2,   \label{tildeF2}
\end{eqnarray}
where we have used the step function (\ref{nstep}) for the distribution function of quasiparticles when integrating over ${\bf p}_1$, ${\bf p}_2$, and ${\bf p}'_1$.
\end{widetext}

For the analysis of the zero sound modes we will need the $p^2$ terms of the interaction function only at ${\bf p}={\bf p}'$. One can easily check that in this case the integral in Eq.(\ref{tildeF2}) is equal to zero. In the integral of Eq.(\ref{tildeF1}) we use the notations:
\begin{equation}   \label{notations}
\frac{{\bf p}+{\bf p}'}{2p_{F0}}={\bf w};\,\,\,\,\,\frac{{\bf p}-{\bf p}'}{2p_{F0}}={\bf b};\,\,\,\,\,\frac{{\bf p}_1+{\bf p}_2}{2p_{F0}}={\bf y}.
\end{equation}
Then we have ${\bf p}_1/p_{F0}={\bf y}-{\bf b};\,({\bf p}_1-{\bf p}')/p_{F0}={\bf y}-{\bf w},$ and equations (\ref{tildeF}-\ref{tildeF2}) are reduced to:
\begin{widetext}
\begin{equation}     
\!\!\!\!F_{mb}({\bf p},{\bf p}')\!\!=\!\tilde F_1({\bf p},{\bf p}')\!\!=\!\frac{\hbar^2}{m}(p_{F0}r_*)^2\!\!\!\int_0^{y_{max}}\!\!\!\!\!\!\!\!\!\frac{d^2y}{by\cos\phi_{yb}}\left\{\!\cos^2\theta_0(2b\!-\!|{\bf y}\!-\!{\bf w}|)
\!+\!\sin^2\theta_0\!\left[\!\frac{(y\cos\phi_y\!-\!w\cos\phi_w)^2}{|{\bf y}\!-\!{\bf w}|}\!-\!2b\cos^2\phi_b\right]\!\right\}^2\!\!\!\!,\!\!\! \label{Fmbinterm}
\end{equation}
where $\phi_y,\,\phi_w,\,\phi_b$ are the angles between the vectors ${\bf y},\,{\bf w},\,{\bf b}$ and the $x$ axis, and $\phi_{yb}$ is the angle between ${\bf y}$ and ${\bf b}$.
The variable $y$ changes from $0$ to $y_{max}=b\cos\phi_{yb}+\sqrt{1-b^2\sin^2\phi_{yb}}$. For ${\bf p}'\rightarrow {\bf p}$ we have $b\rightarrow 0$, and omitting the terms proportional to $b^2$ and higher powers of $b$, we may write $y_{max}=1+b\cos\phi_{yb}$.  
We now recall that $F({\bf p},{\bf p}')=F({\bf p}',{\bf p})$. It is easy to check that $F({\bf p}',{\bf p})$ is given by the same equation (\ref{Fmbinterm}), but with a different sign and $y_{max}=1-b\cos\phi_{yb}$. Then, for ${\bf p}'\rightarrow {\bf p}$ ($b\rightarrow 0$) we have:
\begin{eqnarray}     
&&\frac{F({\bf p},{\bf p}')+F({\bf p}',{\bf p})}{2}=F({\bf p},{\bf p})=\frac{2\hbar^2}{m}(p_{F0}r_*)^2\int_{1-b\cos\phi_{yb}}^{1+b\cos\phi_{yb}}dy\int_0^{2\pi}\frac{d\phi_y}{4b\cos\phi_{yb}}\Big\{\cos^2\theta_0(2b-|{\bf y}-{\bf w}|) \nonumber \\
&&+\sin^2\theta_0\left[\frac{(y\cos\phi_y-w\cos\phi_w)^2}{|{\bf y}-{\bf w}|}-2b\cos^2\phi_b\right]\Big\}^2.\label{Fkk}
\end{eqnarray}
For $b\rightarrow 0$ the result of the integration over $dy$ in Eq.(\ref{Fkk}) is simply obtained by putting $y=1$ in the integrand and multiplying it by $2b\cos\phi_{yb}$. Then, omitting the terms proportional to $b$ and putting $w=1$ we obtain:
\begin{equation}     \label{Fkkinterm}
\!\!\!\!F({\bf p},{\bf p})\!=\!\frac{\hbar^2}{m}(p_{F0}r_*)^2\!\!\!\int_0^{2\pi}\!\!\!\!\!\!d\phi_y\left\{\!2(1\!-\!\cos\phi_{ys})\cos^4\theta_0\!-\!2\sin^2\theta_0\cos^2\theta_0(\cos\phi_y\!-\!\cos\phi_s)^2\!+\!\sin^4\theta_0\frac{(\cos\phi_y\!-\!\cos\phi_s)^4}{2(1\!-\!\cos\phi_{ys})}\right\}\!.\!\!\!
\end{equation}
The integration in Eq.(\ref{Fkkinterm}) is straightforward and it gives:
\begin{equation}    \label{Fmbfin}
F({\bf p},{\bf p})=\frac{4\pi\hbar^2}{m}(p_{F0}r_*)^2\left\{\cos^4\theta_0-\sin^2\theta_0\cos^2\theta_0\left(\frac{1}{2}+\cos^2\phi_p\right)+\sin^4\theta_0\left(\frac{1}{8}+\frac{1}{2}\cos^2\phi_p\right)\right\}.
\end{equation}
For dipoles perpendicular to the plane of their translational motion ($\theta_0=0$) equation (\ref{Fmbfin}) reproduces the result of Ref.~\cite{Lu}.
\end{widetext}.

\section{Dynamical structure factor and zero sound modes}

We now calculate the dynamical structure factor and analyze zero sound modes. Consider a small scalar potential 
\begin{equation}   \label{Phi}
\Phi({\bf r},t)=\Phi({\bf k},\omega)\exp(i{\bf k}{\bf r}-i\omega t) 
\end{equation}
acting on the system via the interaction Hamiltonian
\begin{equation}      \label{He}
\hat H_e=\int\hat\rho({\bf r},t)\Phi({\bf r},t)d^2r,
\end{equation}
where $\hat\rho({\bf r},t)$ is the operator of the particle density. The linear density response function of the system, which is the density-density correlation function, is defined as
\begin{equation}     \label{chi}
\chi({\bf k},\omega)=\frac{d\langle\rho({\bf k},\omega)\rangle}{d\Phi({\bf k},\omega)}\Big|_{\Phi\rightarrow 0},
\end{equation}
with the symbol $\langle...\rangle$ standing for the statistical average. The dynamical structure factor $S({\bf k},\omega)$ is related to the imaginary part of the response function:
\begin{equation}    \label{S}
-\pi[S({\bf k},\omega)-S({\bf k},-\omega)]={\rm Im}\chi({\bf k},\omega).
\end{equation}  

In the collisionless regime, where the frequency $\omega$ of variations of the momentum distribution function greatly exceeds the relaxation rate, the distribution variations $\delta n({\bf p},{\bf r},t)$ are related to deformations of the Fermi surface. Omitting the collisional integral, the kinetic equation in the presence of an external force reads \cite{Pines}:
\begin{equation}           \label{kineq1}
\!\!\!\frac{\partial \delta n}{\partial t}\!+\!\frac{\partial\epsilon({\bf p})}{\partial{\hbar\bf p}}\cdot\frac{\partial \delta n}{\partial \vec{r}}\!-\!\frac{\partial n({\bf p})}{\partial \hbar\vec{p}}\cdot\left\{\!\frac{\partial\delta\epsilon({\bf p})}{\partial \vec{r}}\!+\!\frac{\partial\Phi({\bf r},t)}{\partial {\bf r}}\!\right\}\!=\!0,\!\! 
\end{equation}   
where $n({\bf p})$ is the equilibrium distribution function, $\epsilon({\bf p})$ is the quasiparticle energy at equilibrium, and its variations $\delta\epsilon({\bf p},{\bf r},t)$ are related to the variations of the distribution function through the interaction function of quasiparticles:
\begin{equation}       \label{epsilonF}
\frac{\partial\delta\epsilon({\bf p},{\bf r},t)}{\partial {\bf r}}=\int F({\bf p},{\bf p}')\frac{\partial\delta n({\bf p}',{\bf r},t)}{\partial {\bf r}}\,\frac{d^2p'}{(2\pi)^2}.
\end{equation}
Relying on Eq.(\ref{Phi}) we represent the variations of the distribution function in the form $\delta n({\bf p},{\bf r},t)=\delta n({\bf p},{\bf k},\omega)\exp(i{\bf k}{\bf r}-i\omega t)$ and transform equation (\ref{kineq1}) to
\begin{widetext}
\begin{equation}      \label{kineq2}
({\bf k}{\bf v}_{{\bf p}}-\omega-i\eta)\delta n({\bf p},{\bf k},\omega)-{\bf k}{\bf v}_{{\bf p}}\frac{\partial n({\bf p})}{\partial\epsilon({\bf p})}\left\{\int F({\bf p},{\bf p}')\delta n({\bf p}',{\bf k},\omega)\frac{d^2p'}{(2\pi)^2}\,+\Phi({\bf k},\omega)\right\}=0,
\end{equation}
with $\eta\rightarrow +0$ and ${\bf v}_{{\bf p}}=\partial\epsilon({\bf p})/\partial\hbar{\bf p}$, and we also took into account that $\partial n({\bf p})/\partial\hbar{\bf p}={\bf v}_{{\bf p}}\partial n({\bf p})/\partial\epsilon({\bf p})$.   
\end{widetext}

The distribution variations $\delta n({\bf p},{\bf k},\omega)$ are different from zero only for momenta ${\bf p}$ near the Fermi surface, where ${\bf v}_{{\bf p}}={\bf v}_F({\bf p})$. Putting $\Phi({\bf k},\omega)=0$ in Eq.(\ref{kineq2}) one finds a dispersion relation for the excitation modes. In the limit of interactions tending to zero ($F\rightarrow 0$) we immediately obtain particle-hole modes near the Fermi surface, $\omega={\bf k}{\bf v}_F({\bf p})$. Collective zero sound modes will be obtained below in this section.

We first note that the quantity $\langle\rho({\bf k},\omega)\rangle$ entering equation (\ref{chi}) for the density response function, is given by
\begin{equation}      \label{rho}
\langle\rho({\bf k},\omega)\rangle=\int\delta n({\bf p},{\bf k},\omega)\,\frac{d^2p}{(2\pi)^2},
\end{equation}
and we calculate $\delta n({\bf p},{\bf k},\omega)$. Introducing the function $\tilde\nu(\hat{\bf p})$:
\begin{equation}     \label{nu}
\delta n({\bf p},{\bf k},\omega)=-\frac{{\bf k}{\bf v}_F({\bf p})}{{\bf k}{\bf v}_F({\bf p})-\omega-i\eta}\,\frac{\partial n({\bf p})}{\partial\epsilon({\bf p})}\tilde\nu(\hat{\bf p}),
\end{equation}
equation (\ref{kineq2}) is reduced to
\begin{widetext}
\begin{equation}      \label{kineq3}
\tilde\nu(\hat{\bf p})+\frac{1}{\hbar}\int_0^{2\pi} F(p_F(\phi_p){\bf n},p_F(\phi_{p'}){\bf n}')\frac{{\bf k}{\bf v}_F({\bf p}')}{{\bf k}{\bf v}_F({\bf p}')-\omega-i\eta}\,\frac{p_F(\phi_{p'})}{\hat{\bf p}'{\bf v}_F({\bf p}')}\tilde\nu(\hat{\bf p}')\frac{d\phi_{p'}}{(2\pi)^2}\,+\Phi({\bf k},\omega)=0, 
\end{equation}   
where ${\bf n}$ and ${\bf n}'$ are unit vectors in the directions of ${\bf p}$ and ${\bf p}'$, respectively.  
\end{widetext}

Due to the anisotropy of the Fermi surface, the vectors ${\bf p}$ and ${\bf v}_F({\bf p})$ are not parallel to each other and both $p_F$ and $v_F$ depend on the angle $\phi_p$ between $\hat{\bf p}$ and the $x$ axis. However, the  anisotropy is small and it only leads to a small correction to the term $p_F(\phi_{p'})/\hat{\bf p}'{\bf v}_F({\bf p}')$ in the integrand of Eq.(\ref{kineq3}), so that this term can be put equal to $m/\hbar$. We will also represent the scalar product ${\bf k}{\bf v}_F({\bf p})$ as $kv_F(\phi_p)\cos(\phi_p-\phi_k)$. We thus write Eq.(\ref{kineq3}) as:
\begin{widetext}
\begin{equation}    \label{kineq4}
\nu(\phi_p)-\frac{m}{\hbar^2}\int_0^{2\pi} F(\phi_p,\phi_{p'})\frac{\nu(\phi_{p'})\cos(\phi_{p'}-\phi_k)}{s(\phi_{p'})-\cos(\phi_{p'}-\phi_k)+i\eta}\,\frac{d\phi_{p'}}{4\pi^2}\,+\Phi({\bf k},\omega)=0,
\end{equation}
\end{widetext}
where $s(\phi_p)=\omega/kv_F(\phi_p)$. Assuming $|s-1|\ll 1$ we integrate in Eq.(\ref{kineq4}) singling out the contribution from $\phi_{p'}$ near $\phi_k$ and denoting the rest of the integration as $C(\phi_p)$. This yields:
\begin{equation}     \label{kineq5}
\tilde\nu(\phi_p)-C(\phi_p)+\Phi({\bf k},\omega)-\frac{mF(\phi_p,\phi_k)\tilde\nu(\phi_k)}{2\pi\hbar^2\sqrt{s^2(\phi_k)-1}}=0.
\end{equation}
Substituting the obtained $\tilde\nu(\phi_p)$ into equation (\ref{kineq4}) we find a relation:
\begin{widetext}
\begin{equation}     \label{kineq6}
\tilde\nu(\phi_p)-\frac{mF(\phi_p,\phi_k)[C(\phi_k)-\Phi({\bf k},\omega)]}{2\pi\hbar^2\sqrt{s^2(\phi_k)-1}}-\frac{m^2}{8\pi^3\hbar^4}\int_0^{2\pi}\!\!\!\frac{F(\phi_p,\phi_{p'})F(\phi_{p'},\phi_k)\cos(\phi_{p'}-\phi_k)\tilde\nu(\phi_k)}{\sqrt{s^2(\phi_k)-1}\,[s(\phi_{p'})-\cos(\phi_{p'}-\phi_k)]}d\phi_{p'}\,+\Phi({\bf k},\omega)=0.
\end{equation}
Here we omitted the unimportant contribution of $\phi_{p'}$ away from $\phi_k$ in the integral 
$$\frac{m}{4\pi^2\hbar^2}\int_0^{2\pi}\frac{F(\phi_p,\phi_{p'})[C(\phi_{p'})-\Phi({\bf k},\omega)]\cos(\phi_{p'}-\phi_k)}{s(\phi_{p'})-\cos(\phi_{p'}-\phi_k)}d\phi_{p'}.$$
From equation (\ref{kineq5}) we immediately see that $C(\phi_k)-\Phi({\bf k},\omega)=\tilde\nu(\phi_k)[1-mF(\phi_k,\phi_k)/(2\pi\hbar^2\sqrt{s^2(\phi_k)-1})]$. Then, putting $\phi_p=\phi_k$ in Eq.(\ref{kineq6}) gives:
\begin{equation}     \label{nuphik}
\!\!\!\tilde\nu(\phi_k)\!=\!-\Phi({\bf k},\omega)\!\left[1\!-\!\!\frac{mF(\phi_k,\phi_k)}{2\pi\hbar^2\sqrt{s^2(\phi_k)\!-\!1}}\!+\!\left(\!\frac{mF(\phi_k,\phi_k)}{2\pi\hbar^2\sqrt{s^2(\phi_k)\!-\!1}}\!\right)^2\!\!\!-\!\frac{m^2}{8\pi^3\hbar^4}\!\int_0^{2\pi}\!\!\!\!\!\!\!\frac{F^2(\phi_k,\phi_{p'})\cos(\phi_{p'}\!-\!\phi_k)d\phi_{p'}}{\sqrt{s^2(\phi_k)-1}\,[s(\phi_{p'})-\cos(\phi_{p'}-\phi_k)]}\right]^{-1}\!\!\!\!\!\!\!\!.\!\!\!
\end{equation}
\end{widetext}

The mean field contribution to $F(\phi_k,\phi_k)$ is equal to zero, and hence in the second and third terms in the square brackets in Eq.(\ref{nuphik}) we have to use the many-body interaction function $F(\phi_k,\phi_k)$ given by Eq.(\ref{Fmbfin}). The third term is exactly cancelled by the contribution of $\phi_{p'}$ near $\phi_k$ to the integral in fourth term. For $\phi_{p'}$ away from $\phi_k$ in this integral, we should use the mean field contribution to the interaction function given by Eq.(\ref{F1}), which is linear in $p_{F0}r_*$. This is because the use of quadratic contributions to the interaction function would lead to terms proportional to cubic or higher order powers of $p_{F0}r_*$, which are much smaller than the second term in the square brackets. Representing Eq.(\ref{F1}) on the Fermi surface in the form:
\begin{eqnarray}
&&F_1(\phi_k,\phi_{p'})=4\pi d^2p_{F0}\sin\frac{|\phi_k-\phi_{p'}|}{2}\Big\{P_2(\cos\theta_0) \nonumber \\
&&+\frac{1}{2}\sin^2\theta_0\cos(\phi_k+\phi_{p'})\Big\} \label{F1new}
\end{eqnarray}
and putting $s(\phi_{p'})=1$, the integration of the fourth term in the square brackets in Eq.(\ref{nuphik}) yields:
\begin{widetext}
\begin{equation}         \label{3term}
\frac{m^2}{8\pi^3\hbar^4}\int_0^{2\pi}\frac{F^2(\phi_k,\phi_{p'})\cos(\phi_{p'}-\phi_k)d\phi_{p'}}{\sqrt{s^2(\phi_k)-1}\,[s(\phi_{p'})-\cos(\phi_{p'}-\phi_k)]}=\frac{(p_{F0}r_*)^2P_2(\cos\theta_0)\sin^2\theta_0(2\cos^2\phi_k-1)}{\sqrt{s^2(\phi_k)-1}}.
\end{equation}  
In Eqs.~(\ref{F1new}) and (\ref{3term}) we omitted the anisotropy of the Fermi surface as it leads to corrections which contain higher orders of $p_{F0}r_*$. As a result, equation (\ref{nuphik}) transforms to
\begin{equation}     \label{nuphikfin}
\tilde\nu(\phi_k)=-\Phi({\bf k},\omega)\left[1-\frac{2(p_{F0}r_*)^2\left\{P_2^2(\cos\theta_0)+\frac{1}{8}\sin^4\theta_0\right\}}{\sqrt{s^2(\phi_k)-1}}\right]^{-1}.
\end{equation}
\end{widetext}

The pole of $\tilde\nu(\phi_k)$ corresponds to the solution of Eq.(\ref{kineq4}) with $\Phi({\bf k},\omega)=0$, i.e. to the eigenmodes of zero sound. In the expression for $s_0(\phi_k)$ the anisotropy due to mean field effects and the anisotropy due to the many-body contribution cancel each other, and $s_0$ becomes independent of $\phi_k$:
\begin{equation}     \label{s0}
s_0=1+2(p_{F0}r_*)^4\left\{P_2^2(\cos\theta_0)+\frac{1}{8}\sin^4\theta_0\right\}^2.
\end{equation}
For $\theta_0=0$ equation (\ref{s0}) reproduces the result of Ref.~\cite{Lu}. The zero sound velocity in the dispersion relation $\omega=v_0k$ is:
\begin{equation}     \label{v0}
v_0=v_F(\phi_k)s_0.
\end{equation}
Thus, the zero sound modes exist at any tilting angle $\theta_0$ (for $\theta_0>\theta_c\simeq 0.72$ this is true at temperatures exceeding the critical temperature of the superfluid transition). The anisotropy of the zero sound is practically the same as the anisotropy of the Fermi velocity (omitting corrections to $v_F$ which are $\propto (p_{F0}r_*)^2$). Note that without the many-body contribution to the interaction function we would obtain that the propagation of zero sound is possible only when the result of Eq.(\ref{3term}) is positive. Namely, the zero sound exists for $\theta_0<\arccos(1/\sqrt{3})\simeq 0.96$ and is anisotropic requiring $\phi_k<\pi/4$, or it exists for $\theta_0>0.96$ and $\phi_k>\pi/4$. This is consistent with numerical calculations of Ref.~\cite{Baranov}. As we see, the many-body contribution drastically changes the result.

We now return to Eq.(\ref{nuphikfin}) and use it for obtaining the linear response function on the basis of equations (\ref{chi}), (\ref{rho}), and (\ref{nu}). Integrating in Eq.(\ref{rho}) and dividing the result by $\Phi({\bf k},\omega)$ we obtain:
\begin{widetext}
\begin{equation}     \label{chifinal}
\!\!\!\!\chi({\bf k},\omega)\!=\!\int\frac{d^2p}{(2\pi)^2}\frac{\cos(\phi_p\!-\!\phi_k)}{s(\phi_p)\!-\!\cos(\phi_p-\phi_k)\!+\!i\eta}\,\frac{\partial n({\bf p})}{\partial\epsilon({\bf p})}\frac{\tilde\nu(\phi_p)}{\Phi({\bf k},\omega)}\!=\!-\frac{ms}{2\pi\hbar^2\sqrt{s^2\!-\!1}}\frac{\tilde\nu(\phi_k)}{\Phi({\bf k},\omega)}\!=\!\frac{m}{2\pi\hbar^2}\frac{s}{\sqrt{s^2\!-\!1}\!-\!\sqrt{s_0^2(\phi_k)\!-\!1}},
\end{equation}
\end{widetext}
where $s\equiv s(\phi_k)$. Actually, we should have put $s=1$ in the numerator of Eq.(\ref{chifinal}). However, keeping $s$ in the numerator of the expression for $\chi({\bf k},\omega)$ makes it consistent with the result for a non-interacting 2D gas, which corresponds to $s_0=1$ and is valid for any $s<1$. Thus, equation (\ref{chifinal}) becomes also valid for any $s$ significantly smaller than unity, where the interaction between particles is not important. Relying on equations (\ref{S}) and (\ref{chifinal}) we straightforwardly calculate the dynamical structure factor. For $s<1$ we have:
\begin{equation}    \label{Ssnegative}
S({\bf k},\omega)=\frac{m}{(2\pi\hbar)^2}\,\frac{s\sqrt{1-s^2}}{s_0^2-s^2};\,\,\,\,\,\,s<1.
\end{equation}
For $s>1$ there is only a $\delta$-functional contribution of the zero sound:
\begin{equation}    \label{Sspositive}
S({\bf k},\omega)=\frac{m}{(2\pi\hbar)^2}\,s\sqrt{s_0^2-1}\,\,\delta(s-s_0);\,\,\,\,\,\,s>1.
\end{equation}

The obtained dynamical structure factor is shown in Fig.2. Note that for a non-interacting 2D Fermi gas one obtains a square root singularity in $S$ for $\omega/kv_F\rightarrow 1$. The interaction between particles eliminates this singularity and we have $S$ vanishing as $\sqrt{1-(\omega/kv_F)^2}$ for $\omega/kv_F\rightarrow 1$, which is different from the 3D unpolarized interacting Fermi gas where $S$ vanishes logarithmically for $\omega/kv_F\rightarrow 1$. 

\begin{figure}
\includegraphics[width=9cm]{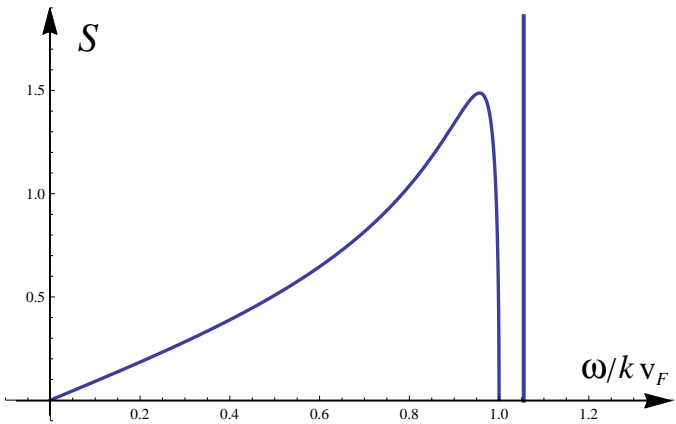}
\caption{(color online) Dynamical structure factor (in units of $m/(2\pi\hbar^2)^2$) as a function of $\omega/kv_F$ for $s_0=1.05$.}
\end{figure}

\section{Relaxation rate of quasiparticles}

The conditions of the collisionless regime, required for the existence of zero sound, are easily achievable in experiments with polar molecules or magnetic atoms. This is seen from the dimensional estimate of the relaxation rate of quasiparticles. At temperatures $T\ll \epsilon_F$ the relaxation of a non-equilibrium distribution of quasiparticles occurs due to binary collisions of quasiparticles which have energies in a narrow interval near the Fermi surface. The width of this interval is $\sim T$ and, hence, the relaxation rate contains a small factor $(T/\epsilon_F)^2$ (see, e.g. \cite{Landau9}). The rate is also proportional to the 2D density $n$ and to the density of states on the Fermi surface, which is $\sim m/\hbar^2$. Using the Fermi golden rule we may write a dimensional estimate for the inverse relaxation time as $\tau^{-1}\sim (g_{eff}^2/\hbar)(m/\hbar^2)n(T/\epsilon_F)^2$, where the quantity $g_{eff}$ is the effective interaction strength. Confining ourselves to the leading part of this quantity, from Eq.~(\ref{F1}) we have $g_{eff}\sim p_Fd^2\sim \hbar^2p_Fr_*/m$. We thus obtain:
\begin{equation}   \label{taurel}
\frac{1}{\tau}\sim\frac{\hbar n}{m}(p_Fr_*)^2\left(\frac{T}{\epsilon_F}\right)^2\sim \frac{mT^2r_*^2}{\hbar^3}.
\end{equation}
Interestingly, for considered temperatures $T\ll \epsilon_F$ the relaxation time $\tau$ is density independent. Excitations with frequencies $\omega\gg 1/\tau$ are in the collisionless regime. Assuming $T\sim 10$nK, for dysprosium atoms which have magnetic moment $10\mu_B$ equivalent to the dipole moment $d\simeq 0.1$ D, we find that $\tau$ is on the level of tenths of a second. The required condition $T\ll\epsilon_F$ is satisfied for $\epsilon_F\gtrsim 50$ nK, which corresponds to $n\gtrsim 3\cdot 10^8$ cm$^{-2}$. In such conditions excitations with frequencies of the order of a Hertz or higher will be in the collisionless regime. 

The occurrence of relaxation of excitations in the collisionless regime is important for understanding the visibility of zero sound.
The dynamical structure factor $S({\bf k},\omega)$ characterizes the scattering process in which the momentum ${\bf k}$ and energy $\hbar\omega$ are transferred to the system. The visibility of the zero sound peak in the structure factor can be smeared out by the fact that it is very close to the particle-hole continuum (see Fig.2). The related distance is $\sim (p_{F0}r_*)^4$ in units of $\omega/v_Fk$. In order to make sure that this is not the case one has to find the actual height and width of the zero sound peak. Also, one can think of observing the oscillations of the cloud induced by small modulations of the density, with a time delay after switching off the driving force perturbing the density. In both cases the picture is determined by the damping of zero sound and quasiparticles excitations.  

Thus, we should compare the relaxation rate of quasiparticles near the Fermi surface with the damping rate of zero sound. First, we calculate the rate of relaxation of a quasiparticle near the Fermi surface, with a given energy $\epsilon({\bf p})\ll\epsilon_{F0}$ at $T\rightarrow 0$. The relaxation mechanism involves the interaction of this quasiparticle with the filled Fermi sphere, which annihilates the quasiparticle, creates a hole with momentum ${\bf p}_1$ (annihilates a particle with momentum ${\bf p}_1$ inside the Fermi sphere), and creates quasiparticles with momenta ${\bf p}_2$ and ${\bf p}_3$. As the relaxation rate $\tau^{-1}$ is small, we use the first order perturbation theory (Fermi golden rule) relying on the interaction Hamiltonian $\hat H_{int}$ given by the second term of Eq.(\ref{H}). We then have:
\begin{eqnarray}        
\frac{1}{\tau}&=&\frac{2\pi}{\hbar}\sum_{{\bf p}_1,{\bf p}_2,{\bf p}_3}|\langle\hat a_{{\bf p}_1}^{\dagger}\hat a_{{\bf p}_2}\hat a_{{\bf p}_3}\hat H_{int}\hat a_{{\bf p}}^{\dagger}\rangle|^2\nonumber \\
&\times&\delta(\epsilon({\bf p})+\epsilon({\bf p}_1)-\epsilon({\bf p}_2)-\epsilon({\bf p}_3)),\label{tau1}
\end{eqnarray}    
where the symbol $\langle...\rangle$ stands for the average over the equilibrium state, and
\begin{eqnarray}
\hat H_{int}=\frac{1}{2S}\sum_{{\bf p}_1,{\bf p}_2,{\bf p}_3}U({\bf p}_1-{\bf p}_3)\hat a_{{\bf p}_3}^{\dagger}\hat a_{{\bf p}_4}^{\dagger}\hat a_{{\bf p}_2}\hat a_{{\bf p}_1}. \nonumber
\end{eqnarray}
Using the Wick theorem and the relations $\langle\hat a_{{\bf p}_i}^{\dagger}\hat a_{{\bf p}'_i}\rangle=n({\bf p}_i)\delta_{{\bf p}_i{\bf p}'_i},\,\,\langle\hat a_{{\bf p}_i}
\hat a_{{\bf p}'_i}^{\dagger}\rangle=(1-n({\bf p}_i))\delta_{{\bf p}_i{\bf p}'_i}$, we reduce equation (\ref{tau1}) to
\begin{widetext}           
\begin{eqnarray}      
\frac{1}{\tau}&=&\frac{2\pi}{\hbar}\int_0^{\infty}\frac{p_1dp_1}{(2\pi)^2}\int_0^{\infty}\frac{k_2dk_2}{(2\pi)^2}\int_0^{2\pi}d\phi_1\int_0^{2\pi}d\phi_2\,n({\bf p}_1)(1-n({\bf p}_2))(1-n({\bf p}_3))[U({\bf p}-{\bf p}_2)-U({\bf p}_1-{\bf p}_2)]^2  \nonumber \\
&\times&\delta(\epsilon({\bf p})+\epsilon({\bf p}_1)-\epsilon({\bf p}_2)-\epsilon({\bf p}_3))\delta_{{\bf p}+{\bf p}_1,\,{\bf p}_2+{\bf p}_3},\label{tau2}
\end{eqnarray}
where $\phi_1$ ($\phi_2$) are the angles between the vectors ${\bf p}_1$ (${\bf p}_2$) and the $x$ axis, and the presence of the Kronecker symbol $\delta_{{\bf p}+{\bf p}_1,\,{\bf p}_2+{\bf p}_3}$ reflects the momentum conserva‭tion law.
\end{widetext}

We now omit the small anisotropy of the Fermi surface in the $\delta$-function and occupation numbers. Since the anisotropy is omitted in all derivations below, we will use the notations $v_F$ and $p_F$ for the Fermi velocity $v_F$ and Fermi momentum $p_{F}$. As all involved quasiparticle states are near the Fermi surface, we represent the energies of these states in the form 
$\epsilon({\bf p}_i)=\hbar v_{F}q_i$, where $q_i=p_i-p_{F}$ and $|q_i|\ll p_{F}$. The particle that undergoes the relaxation is certainly above the Fermi surface, and $q=p-p_{F}>0$. We first integrate in Eq.(\ref{tau2}) over the angles‎ $\phi_2$ and $\phi_1$. The dependence of the integrand on these angles is contained in the $\delta$-function and in the Fourier transforms $U({\bf p}-{\bf p}_2)$ and $U({\bf p}_1-{\bf p}_2)$. From the energy and momentum conservation laws we have $p_3=p+p_1-p_2=|{\bf p}+{\bf p}_1-{\bf p}_2|$, which gives a relation:
\begin{eqnarray}   
&&pp_1[(1-\cos(\phi_1-\phi)]=pp_2[1-\cos(\phi_2-\phi)] \nonumber \\
&&+p_1p_2[1-\cos(\phi_2-\phi_1)],   \label{phi21}
\end{eqnarray}
with $\phi$ being an angle between the vector ${\bf p}$ and $x$ axis. 
It is convenient to represent Eq.(\ref{phi21}) in the form:
\begin{widetext}
\begin{equation}   \label{phi21tilde}
\sin(\phi_2-\phi_1-\tilde\phi)=\frac{pp_1[1-\cos(\phi_1-\phi)]-(p_2p_1+pp_2)}{\sqrt{[p_2p_1+pp_2\cos(\phi_1-\phi)]^2+p^2p_2^2\sin^2(\phi_1-\phi)}},
\end{equation}
where
\begin{eqnarray}
\sin\tilde\phi=\frac{p_2p_1+pp_2\cos(\phi_1-\phi)}{\sqrt{[p_2p_1+pp_2\cos(\phi_1-\phi)]^2+p^2p_2^2\sin^2(\phi_1-\phi)}};\,\,\,\,\,\,\,\cos\tilde\phi=\frac{pp_2\sin(\phi_1-\phi)}{\sqrt{[p_2p_1+pp_2\cos(\phi_1-\phi)]^2+p^2p_2^2\sin^2(\phi_1-\phi)}}. \nonumber
\end{eqnarray}
For the derivative $dp_3/d\phi_2$ we have:
\begin{eqnarray}     
\!\!\!\!\!\left|\frac{dp_3}{d\phi_2}\right|\!\!=\!\frac{\sqrt{[p_2p_1\!\!+\!pp_2\!\cos(\phi_1\!\!-\!\phi)]^2\!+\!p^2p_2^2\sin^2(\!\phi_1\!-\!\phi)}}{p_3}|\!\cos(\phi_2\!-\!\phi_1\!-\!\tilde\phi)\!|\!=\!2p_{F}\!\left|\sin\!\left(\!\frac{\phi_1\!-\!\phi}{2}\!\right)\!\right|\!\sqrt{\!\cos^2\!\left(\!\frac{\phi_1\!-\!\phi}{2}\!\right)\!+\!\!(q\!-\!q_2)(q_2\!-\!q_1\!)/p_{F}^2}. \nonumber
\end{eqnarray}
In the Fourier transforms $U({\bf p}-{\bf p}_2)$ and $U({\bf p}_1-{\bf p}_2)$ we may put $|{\bf p}|=|{\bf p}_1|=|{\bf p}_2|=p_{F}$. Then equation (\ref{phi21}) gives either $\phi_2=\phi_1$ or $\phi_2=\phi$. In both cases, using Eq.(\ref{F1}) we obtain:
\begin{eqnarray}
U({\bf p}-{\bf p}_2)-U({\bf p}_1-{\bf p}_2)=\pm 4\pi d^2p_{F}\left|\sin\left(\frac{\phi_1-\phi}{2}\right)\right|\left\{\cos^2\theta_0-\sin^2\theta_0\sin^2\left(\frac{\phi_1+\phi}{2}\right)\right\}. \label{pm}
\end{eqnarray}
Integrating over $\phi_2$ in Eq.(\ref{tau2}) we then have:
\begin{equation}     \label{I1}
\int_0^{2\pi}d\phi_1\int_0^{2\pi}d\phi_2[U({\bf p}-{\bf p}_2)-U({\bf p}_1-{\bf p}_2)]^2\delta(\epsilon({\bf p})+\epsilon({\bf p}_1)-\epsilon({\bf p}_2)-\epsilon({\bf p}_3))\delta_{{\bf p}+{\bf p}_1,\,{\bf p}_2+{\bf p}_3}=\frac{16\pi^2md^4}{\hbar^2}I(q_1,q_2),
\end{equation}
where the quantity $I(q_1,q_2)$ also depends on $q$, $\theta_0$, $\phi$ and is given by
\begin{equation} \label{I2}          
I(q_1,q_2)=\int_0^{2\pi}d\phi_1\frac{\left|\sin\left(\frac{\phi_1-\phi}{2}\right)\right|\left[\cos^2\theta_0-\sin^2\theta_0\sin^2\left(\frac{\phi_1+\phi}{2}\right)\right]^2}{\sqrt{\!\cos^2\!\left(\!\frac{\phi_1\!-\!\phi}{2}\!\right)\!+\!\!(q\!-\!q_2)(q_2\!-\!q_1\!)/p_{F}^2}}.
\end{equation}
\end{widetext}
Formally, when integrating over $\phi_1$ one should put a constraint $\cos^2[(\phi_1-\phi)/2]\geq (q_2-q)(q_2-q_1)/p_{F}^2$ in order to satisfy the inequality $\sin^2(\phi_2-\phi_1-\tilde\phi)\leq 1$. However, at $T=0$ the created particles are above the Fermi surface, so that $q_2>0$ and $q_3=q+q_1-q_2>0$. The annihilated particle is below the Fermi surface and, hence, $q_1<0$. We thus have $q>q_2$ and $q_2>q_1$, and the inequality $\cos^2[(\phi_1-\phi)/2]\geq (q_2-q)(q_2-q_1)/p_{F}^2$ is satisfied for any $\phi_1$. Putting $n({\bf p}_1)=1,\,n({\bf p}_2)=n({\bf p}_3)=0$ in Eq.(\ref{tau2}) and writing $p_idp_i=p_{F}dq_i$ we set the following limits of integration over $q_1$ and $q_2$:   
\begin{eqnarray}
-q\leq &q_1&\leq 0;   \label{q1limits} \\
0\leq &q_2&\leq q+q_1.  \label{q2limits}
\end{eqnarray}

Equations (\ref{I1}) and (\ref{I2}) describe the contributions which correspond to $\phi_2$ close to $\phi_1$ or to $\phi$. Another contribution comes from $\phi_1$ close to $\phi+\pi$. In this case we may put $\phi_1=\phi+\pi$ in the Fourier transform $U({\bf p}_1-{\bf p}_2)$, which gives:
\begin{widetext}
\begin{eqnarray}      
&&|U({\bf p}-{\bf p}_2)-U({\bf p}_1-{\bf p}_2)|=4\pi d^2p_{F}\Big[\left|\sin\left(\frac{\phi_1-\phi}{2}\right)\right|\left\{\cos^2\theta_0-\sin^2\theta_0\sin^2\left(\frac{\phi_1+\phi}{2}\right)\right\}    \nonumber \\
&&-\left|\cos\left(\frac{\phi_1-\phi}{2}\right)\right|\left\{\cos^2\theta_0-\sin^2\theta_0\cos^2\left(\frac{\phi_1+\phi}{2}\right)\right\}\Big]. \label{pm2} 
\end{eqnarray}
Performing similar calculations as above we obtain Eq.(\ref{I1}) in which the function $I(q_1,q_2)$ is replaced with $\tilde I(q_1,q_2)$ given by (see \cite{shortnote}):
\begin{eqnarray}
&&\tilde I(q_1,q_2)=\int_0^{2\pi}d\phi_2\Big\{\frac{\left|\sin\left(\frac{\phi_2-\phi}{2}\right)\right|\left[\cos^2\theta_0-\sin^2\theta_0\sin^2\left(\frac{\phi_2+\phi}{2}\right)\right]^2}{2\sqrt{\!\cos^2\left(\!\frac{\phi_2\!-\!\phi}{2}\!\right)-2(q_2\!-\!q_1\!)/p_{F}}}-\left[\cos^2\theta_0-\sin^2\theta_0\sin^2\left(\frac{\phi_2+\phi}{2}\right)\right] \nonumber    \\
&&\times\left[\cos^2\theta_0-\sin^2\theta_0\cos^2\left(\frac{\phi_2+\phi}{2}\right)\right] 
+\frac{2\left|\sin\left(\frac{\phi_2-\phi}{2}\right)\right|\left|\cos\left(\frac{\phi_2-\phi}{2}\right)\right|}{4\sin^2\left(\frac{\phi_2-\phi}{2}\right)+(q-q_2)^2/p_{F}^2}\left[\cos^2\theta_0-\sin^2\theta_0\cos^2\left(\frac{\phi_2+\phi}{2}\right)\right]^2\Big\}, \label{tildeI}
\end{eqnarray}
where we keep only leading powers of $q,q_1,q_2$. Note that when integrating the first term in the curly brackets we have a constraint that the argument of the square root in the denominator is positive.
\end{widetext}

With both contributions taken into account, equation (\ref{tau2}) reduces to
\begin{equation}    \label{tau3}
\!\!\frac{1}{\tau}\!=\!\frac{2\hbar}{\pi m}(p_{F}r_*)^2\!\!\!\int_{-q}^0\!\!\!dq_1\!\!\int_0^{q_1+q}\!\!dq_2(I(q_1,q_2)+\tilde I(q_1,q_2)).\!\!
\end{equation}
The integrals (\ref{I2}) and (\ref{tildeI}) take the forms
\begin{widetext}
\begin{eqnarray}
\!\!\!\!&&I(q_1,q_2)=4\int_0^1\frac{dx}{\sqrt{x^2+(q-q_2)(q_2-q_1)/p_{F}^2}}\{(\cos^2\theta_0-\sin^2\theta_0\cos^2\phi)^2+2x^2\sin^2\theta_0\cos^2\theta_0(2\cos^2\phi-1) \nonumber \\
\!\!\!\!&&+\sin^4\theta_0[x^4(\sin^4\phi+\cos^4\phi-6\sin^2\phi\cos^2\phi)+x^2(6\cos^2\phi\sin^2\phi-2\cos^4\phi)]\} \nonumber \\
\!\!\!\!&&=2(\cos^2\theta_0-\sin^2\theta_0\cos^2\phi)^2\ln\left(\frac{4p_{F}^2}{(q\!-\!q_2)(q_2-q_1)}\right)+4{\cal F}(\theta_0,\phi);\!\!\!\! \label{tildel3}
\end{eqnarray}
\begin{eqnarray}
\!\!\!\!&&\tilde I(q_1,q_2)=2\int_0^1 \!\!dx\!\left[\frac{1}{\sqrt{x^2+2(q_2-q_1)/p_F}}+\frac{4x}{4x^2+(q-q_2)^2/p_{F}^2}\right]\!\{(\cos^2\theta_0-\sin^2\theta_0\cos^2\phi)^2+2x^2\sin^2\theta_0\cos^2\theta_0(2\cos^2\phi-1) \nonumber \\
\!\!\!\!&&+\sin^4\theta_0[x^4(\sin^4\phi+\cos^4\phi-6\sin^2\phi\cos^2\phi)+x^2(6\cos^2\phi\sin^2\phi-2\cos^4\phi)]\}-2\pi\left(\cos^4\theta_0-\cos^2\theta_0\sin^2\theta_0+\frac{1}{8}\sin^4\theta_0\right)
\nonumber   \\ 
\!\!\!\!&&=(\cos^2\theta_0-\sin^2\theta_0\cos^2\phi)^2\ln\left(\frac{8p_{F}^3}{(q\!-\!q_2)^2(q_2-q_1)}\right)+4\tilde{\cal F}(\theta_0,\phi), \!\!\!\!\!\!\label{tildel4}
\end{eqnarray}
where we omitted higher powers of $q$, $q_1$, and $q_2$ and introduced the functions: 
\begin{eqnarray}    
&&{\cal F}(\theta_0,\phi)=\sin^2\theta_0\cos^2\theta_0(2\cos^2\phi-1)+\sin^4\theta_0\left(\frac{1}{4}+\cos^2\phi-2\cos^4\phi\right), \label{calF}  \\
&&\tilde{\cal F}(\theta_0,\phi)={\cal F}(\theta_0,\phi)-\frac{\pi}{2}\left(\cos^4\theta_0-\cos^2\theta_0\sin^2\theta_0+\frac{1}{8}\sin^4\theta_0\right). \label{caltildeF}
\end{eqnarray}
Substituting the results of Eqs.~(\ref{tildel3}) and (\ref{tildel4}) into equation (\ref{tau3}) and integrating over $q_1$ and $q_2$ we obtain:
\begin{equation}        \label{tau4}
\frac{1}{\tau}=\frac{4\hbar}{\pi m}(p_{F}r_*)^2q^2\left\{\left(\frac{3}{4}\ln\frac{p_{F}^2}{q^2}+\frac{3}{4}+\frac{3}{2}\ln{2}\right)(\cos^2\theta_0-\sin^2\theta_0\cos^2\phi)^2+{\cal F}(\theta_0,\phi)+\tilde{\cal F}(\theta_0,\phi)\right\}.
\end{equation}
\end{widetext}
Recalling that the quasiparticle energy is $\epsilon(q)=\hbar v_{F}q$ and $p^2_{F}=4\pi n$, we represent Eq.(\ref{tau4}) in the form:
\begin{equation}    \label{taufin}
\frac{1}{\tau}=\frac{6\hbar}{m}n(p_{F}r_*)^2\left(\frac{\epsilon(q)}{\epsilon_{F}}\right)^2A(q,\theta_0,\phi),
\end{equation}
where
\begin{eqnarray}  
&&A=(\cos^2\theta_0-\sin^2\theta_0\cos^2\phi)^2\ln\left[\frac{4e^{1/2}\epsilon_{F}}{\epsilon(q)}\right] \nonumber \\
&&+\frac{2}{3}\left({\cal F}(\theta_0,\phi)+\tilde{\cal F}(\theta_0,\phi)\right).\label{A} 
\end{eqnarray}
The dependence $\tau^{-1}\propto q^2$ is generic for Fermi liquids \cite{Landau9}, and the appearance of the logarithmic factor in Eq.(\ref{taufin}) is due to the 2D geometry of the system. 

The obtained relaxation rate strongly depends on the tilting angle $\theta_0$ and on the angle $\phi$ of the quasiparticle wavevector ${\bf p}$ with respect to the tilting direction. The rate reaches maximum when the dipoles are perpendicular to the plane of their translational motion ($\theta_0=0$). In this case 
$$A_{max}=\ln\left[\frac{4e^{1/2}\epsilon_{F}}{\epsilon(q)}\right].$$
The minimum value of $\tau^{-1}$ is achieved for the dipoles lying in the plane of their translational motion ($\theta_0=\pi/2$) at the angle $\phi$ equal to $\pi/2$. We then have
$$A_{min}=\frac{1}{3}-\frac{\pi}{24}\simeq 0.2.$$

The absolute value of the relaxation time of an excitation of a given frequency (in units of the Fermi energy) at a given density, strongly depends on a particular system. For example, in the case of dysprosium atoms ($d\simeq 0.1$D and $r_*\simeq 25$ nm) at a density $n\sim 10^9$ cm$^{-2}$ we have the Fermi energy approaching $200$ nK (5 kHz), and equation (\ref{taufin}) gives the relaxation time $\tau$ of the order of a second or higher for the excitation energy of $10^{-2}\epsilon_F$ (50 Hz). At the same time for NaK molecules, selecting the electric field that provides $d\simeq 0.4$D ($r_*\simeq 100$ nm), for $\epsilon(q)\simeq 10^{-2}\epsilon_F$ (which is 150 Hz as we now have $\epsilon_F\simeq 15$ kHz) we obtain $\tau\approx 20$ ms at the same density of $10^9$ cm$^{-2}$ and $\theta_0=0$. 

\section{Damping of zero sound}

The calculation of the damping rate of zero sound modes is more involved. It has to include the zero sound through the non-equilibrium character of the distribution function. The discussion of this topic has been initiated by Landau \cite{Landau} who assumed that the transition probability for the scattering of quasiparticles with given momenta in the wave of zero sound is the same at temperatures $T\gg\hbar\omega$, where $\omega$ is the frequency of the zero sound, and at $T=0$. He then established a relation between the damping of zero sound at $T\gg\hbar\omega$ and at zero temperature \cite{Landau,Abrikosov1958}. In a later stage, theoretical studies of the attenuation of zero sound in liquid $^3$He were based on microscopic considerations \cite{Eliashberg,Pethick}. 

Following the idea of Landau we first consider the attenuation of zero sound at temperatures $T\gg\hbar\omega$ and start with the kinetic equation (\ref{kineq1}) in which we include the colliasional integral ${\cal I}(n)$ and put the external potential $\Phi\rightarrow 0$:
\begin{equation}    \label{kineqz1}
\frac{\partial \delta n}{\partial t}+\frac{\partial\epsilon({\bf p})}{\partial{\hbar\bf p}}\cdot\frac{\partial \delta n}{\partial \vec{r}}-\frac{\partial n({\bf p})}{\partial \hbar\vec{p}}\cdot\frac{\partial\delta\epsilon({\bf p})}{\partial \vec{r}}={\cal I}(n), 
\end{equation}
where $n({\bf p})$ is the equilibrium distribution function, $\delta n({\bf p},{\bf r},t)$ is the deviation of the distribution function from the equilibrium value, and variations of the quasiparticle energy are expressed through $\delta n$ by Eq.(\ref{epsilonF}).  In  the presence of zero sound, variations of the distribution function follow from Eq.(\ref{nu}). Omitting the small anisotropy of the Fermi surface, $\delta n$ can be written as (see Eq.(\ref{nu})):
\begin{equation}     \label{deltanct}
\delta n({\bf p},{\bf r},t)=-\frac{\partial n({\bf p})}{\partial \epsilon({\bf p})}\nu(\phi)\exp\{i{\bf kr}-i\omega t\},
\end{equation}
where the function $\nu(\phi)$ has a sharp peak for $\phi\rightarrow\phi_k$, with $\phi\equiv\phi_p$ and $\phi_k$ being the angle between the wave vector of the zero sound ${\bf k}$ and the tilting direction. 

At temperatures $T\gg\hbar\omega$ one may omit the frequency and momentum of the zero sound in the energy and momentum conservation laws. Then the collisional integral reduces to the form \cite{Landau10}:
\begin{widetext}
\begin{equation}   \label{In1}
{\cal I}(n)=\frac{1}{T}\int Wn(\epsilon(q))n_1(1-n_2)(1-n_3)(\zeta_2+\zeta_3-\zeta_1-\zeta)\delta(\epsilon(q)+\epsilon_1-\epsilon_2-\epsilon_3)\frac{d^2p_1d^2p_2}{(2\pi)^4},
\end{equation}
where $\epsilon_i=\epsilon(q_i)$, $n_i=n(\epsilon(q_i))$, and $\zeta_i=\nu(\phi_i)+(m/2\pi\hbar^2)\int\nu(\phi_i')F(\phi_i,\phi_i')d\phi_i'$. The momentum conservation law reads ${\bf p}+{\bf p}_1={\bf p}_2+{\bf p}_3$.
\end{widetext}
The quantity $W$ is given by
\begin{equation}    \label{W}
W=\frac{2\pi}{\hbar}[U({\bf p}-{\bf p}_2)-U({\bf p}_1-{\bf p}_2)]^2,
\end{equation}
and the notations are the same as in Section V.

The functions $\zeta_i$ are taken on the Fermi surface, and we can do the same with respect to $U({\bf p}-{\bf p}_2)$ and $U({\bf p}_1-{\bf p}_2)$. The only way to satisfy the momentum conservation on the Fermi surface and get a non-zero quantity $[\zeta(\phi)+\zeta(\phi_1)-\zeta(\phi_2)-\zeta(\phi_3)]$ is to put $\phi_1=\phi+\pi$ (and, hence, $\phi_3=\phi_2+\pi$). We then have $W(\phi_2,\phi)$ following from equation (\ref{W}) with $[U({\bf p}-{\bf p}_2)-U({\bf p}_1-{\bf p}_2)]$ from Eq.(\ref{pm2}), and the collisional integral becomes:
\begin{widetext}     
\begin{equation}   \label{In2}
{\cal I}(n)=\frac{m}{\hbar^2T}\int_0^{2\pi}\frac{d\phi_2}{(2\pi)^4}\int_{-\infty}^{\infty}dq_1\int_{-\infty}^{\infty}dq_2 \frac{W(\phi_2,\phi)}{|\sin(\phi_2-\phi)|}[\zeta(\phi_2)+\zeta(\phi_2+\pi)-\zeta(\phi)-\zeta(\phi+\pi)]n(\epsilon(q))n_1(1-n_2)(1-n_3).
\end{equation}
Assuming $\epsilon_q\ll T$ and using the finite temperature Fermi-Dirac distribution for $n_1$, $n_2$, and $n_3$ we then obtain:
\begin{equation}      \label{In3}
{\cal I}(n)=\frac{\pi^2mT^2}{2(\hbar^2 v_{F})^2}\int_0^{2\pi}\frac{d\phi_2}{(2\pi)^4}\frac{W(\phi_2,\phi)}{|\sin(\phi_2-\phi)|}[\zeta(\phi)+\zeta(\phi+\pi)-\zeta(\phi_2)-\zeta(\phi_2+\pi)]\frac{\partial n(\epsilon(q))}{\partial\epsilon(q)}.
\end{equation}
\end{widetext}

We now set
\begin{equation}    \label{nunu}
\nu(\phi)=\frac{{\bar \nu}(\phi)}{s_0(\phi_k)-\cos(\phi-\phi_k)},
\end{equation}
where ${\bar \nu}(\phi)$ is a smooth function, and $s_0$ is given by Eq.(\ref{s0}). To zero order in $p_Fr_*$ we omit the second term in the expression for $\zeta(\phi_i)$ and then obtain:
\begin{widetext}
\begin{equation}
\int_0^{2\pi}\!\!\frac{\pi^2W(\phi_2,\phi)}{2|\sin(\phi_2-\phi)|}\frac{d\phi_2}{(2\pi)^4}[\zeta(\phi)+\zeta(\phi+\pi)-\zeta(\phi_2)-\zeta(\phi_2+\pi)]\!=\!\frac{\pi\hbar^3}{m^2}(p_{F}r_*)^2B(\phi),
\label{A1}
\end{equation}
where
\begin{eqnarray}   
\!\!\!\!\!\!&&B(\phi)=\!\!\int_0^{2\pi}\!\!d\phi_2\!\!\left(\frac{{\bar \nu}(\phi)}{s_0-\cos(\phi-\phi_k)}+\frac{{\bar \nu}(\phi+\pi)}{s_0+\cos(\phi-\phi_k)}-\frac{{\bar \nu}(\phi_2)}{s_0-\cos(\phi_2-\phi_k)}-\frac{{\bar \nu}(\phi_2+\pi)}{s_0+\cos(\phi_2-\phi_k)}\!\right)   \nonumber  \\
\!\!\!\!\!\!&&\!\times \left\{\left|\sin\!\left(\!\frac{\phi_2\!-\!\phi}{2}\!\right)\right|\left[\cos^2\theta_0\!-\sin^2\theta_0\sin^2\!\left(\!\frac{\phi_2\!+\!\phi}{2}\!\right)\!\right]\!-\!\left|\cos\left(\frac{\!\phi_2\!-\!\phi}{2}\!\right)\right|\left[\cos^2\theta_0\!-\sin^2\theta_0\cos^2\!\left(\!\frac{\!\phi_2\!+\!\phi}{2}\!\right)\right]\right\}^2\!\!\!\frac{d\phi_2}{|\sin(\phi_2\!-\!\phi)|}.   \label{B}      
\end{eqnarray}
To zero order in $p_Fr_*$ we may put all ${\bar \nu}$ functions in Eq.(\ref{B}) equal to ${\bar \nu}(\phi_k)$. This, in particular, yields:
\begin{equation}      \label{Bphik}
B(\phi_k)=\frac{4{\bar \nu}(\phi_k)}{s_0^2-1}\left\{(\cos^2\theta_0-\sin^2\theta_0\cos^2\phi_k)^2\ln\left[\frac{s_0+1}{s_0-1}\right]+2\tilde{\cal F}(\theta_0,\phi_k)\right\},
\end{equation}
where the function $\tilde{\cal F}(\theta_0,\phi_k)$ has been introduced in Eq.(\ref{caltildeF}).
\end{widetext}
Using Eq.(\ref{A1}) the collisional integral (\ref{In3}) reduces to
\begin{equation}    \label{In4}
{\cal I}(n)=\frac{1}{\tau_T}\frac{\partial n}{\partial \epsilon(q)}B(\phi),
\end{equation}
with
\begin{equation}    \label{tauT}
\frac{1}{\tau_T}=\frac{\pi T^2}{2\hbar\epsilon_F}(p_Fr_*)^2,
\end{equation}
and making use of Eqs.~(\ref{deltanct}) and (\ref{In4}) the kinetic equation (\ref{kineqz1}) takes the form:
\begin{widetext}
\begin{equation}     \label{kineqz2}
[\omega-kv_F\cos(\phi-\phi_k)]\nu(\phi)-\frac{mkv_F}{4\pi^2\hbar^2}\cos(\phi-\phi_k)\int_0^{2\pi}d\phi' F(\phi,\phi')\nu(\phi')=-\frac{iB(\phi)}{\tau_T}.
\end{equation}
\end{widetext}

In the presence of damping the zero sound, frequency $\omega$ is complex for real $k$. We will use the notation $\omega/kv_F=s$, where the real part of $s$ is equal to $s_0$ and the imaginary part is related to the attenuation of zero sound. We also assume that the damping rate is much smaller than the shift of the frequency $\omega$ from $kv_F$, given by $kv_F(s_0-1)$. This means that $(s_0-1)$ greatly exceeds the imaginary part of $s$. We thus may first proceed with Eq.(\ref{kineqz2}) in the same way as we did in Section IV (see equations (\ref{kineq3}) - (\ref{kineq6})) and represent (\ref{kineqz2}) in the form similar to Eq.(\ref{kineq6}). The difference is that now we replace $\Phi({\bf k},\omega)$ by the term $iB(\phi)/\tau_T$. We have:
\begin{widetext}
\begin{equation}      \label{kineqz3}
\!\!\!\!\!\tilde\nu(\phi)\!-\!\frac{mF(\phi,\phi_k)\tilde\nu(\phi_k)}{2\pi\hbar^2\sqrt{s^2(\phi_k)\!-\!1}}\!+\!\frac{m^2F(\phi_k,\phi_k)F(\phi,\phi_k)\tilde\nu(\phi_k)}{4\pi^2\hbar^4(s^2(\phi_k)\!-\!1)}\!-\!\frac{m^2}{8\pi^3\hbar^4}\!\!\!\int_0^{2\pi}\!\!\frac{F(\phi,\phi')F(\phi',\phi_k)\tilde\nu(\phi_k)\cos(\phi'\!-\!\phi_k)d\phi'}{\sqrt{s^2(\phi_k)-1}\,[s(\phi')-\cos(\phi'\!-\phi_k)]}\!=\frac{-iB(\phi)}{kv_F\tau_T},\!\!\!\!\!\!\!\!\!\!  
\end{equation}
where the function $\tilde\nu$ has been introduced in Eq.(\ref{nu}) and it is related to $\nu$ as $\nu(\phi)=\tilde\nu(\phi)\cos(\phi-\phi_k)/(s-\cos(\phi-\phi_k))$. The contribution of $\phi'$ close to $\phi_k$ in the integral over $d\phi'$ in the last term of the lhs of Eq.(\ref{kineqz3}) and the third term of the lhs
cancel each other, and Eq.(\ref{kineqz3}) reduces to
\begin{equation}       \label{kineqz4}
\tilde\nu(\phi)-\frac{mF(\phi,\phi_k)\tilde\nu(\phi_k)}{2\pi\hbar^2\sqrt{s^2(\phi_k)\!-\!1}}\!-\!\frac{m^2}{8\pi^3\hbar^4}\!\int_0^{2\pi}\!\frac{F_1(\phi,\phi')F_1(\phi',\phi_k)\tilde\nu(\phi_k)\cos(\phi'\!-\!\phi_k)d\phi'}{\sqrt{s^2(\phi_k)-1}\,[s(\phi')-\cos(\phi'-\phi_k)]}\!=\frac{-iB(\phi)}{kv_F\tau_T},\!\!\!\!\!\!\!\!\!\!  
\end{equation}
with the mean-field interaction function $F_1$ given by Eq.(\ref{F1new}).
\end{widetext}

We now take the limit $\phi\rightarrow\phi_k$ and note that then the lhs of Eq.(\ref{kineqz4}) can be conveniently expressed in terms of $s_0$ and $s$, which leads to the relation:
\begin{equation}     \label{kineqz5}
\tilde\nu(\phi_k)\left(1-\sqrt{\frac{s_0^2-1}{s^2-1}}\right)=-\frac{iB(\phi_k)}{kv_F\tau_T}.
\end{equation}
In equation (\ref{Bphik}) for $B(\phi_k)$ we may replace ${\bar \nu}(\phi_k)$ 
with $\tilde\nu(\phi_k)$ and thus obtain from Eq.(\ref{kineqz5}):
\begin{equation}    \label{Ims}
{\rm Im}s=-\frac{8}{kv_F\tau_T}D(\theta_0,\phi_k),
\end{equation}
with
\begin{eqnarray}    
D(\theta_0,\phi_k)&=&\frac{1}{2}(\cos^2\theta_0-\sin^2\theta_0\cos^2\phi_k)^2\ln\left[
\frac{s_0+1}{s_0-1}\right]   \nonumber \\
&+&\tilde{\cal F}(\theta_0,\phi_k). \label{D}
\end{eqnarray}
Writing the zero sound frequency as $\omega=s_0kv_F-i/2\tau_{0T}$, for the damping rate $\tau_{0T}^{-1}$ we find:
\begin{equation}   \label{tau0T}
\frac{1}{\tau_{0T}}=\frac{16D(\theta_0,\phi_k)}{\tau_T}.
\end{equation}

We now proceed in the same way as has been done in Ref. \cite{Pethick} for the attenuation of zero sounf in $^3$He and as described in \cite{Landau10}. In the regime where the zero sound frequency $\omega$ is comparable with $T$ or exceeds it, the reduction of the number of the zero sound quanta per unit time due to quasiparticle collisions is given by
\begin{widetext}
\begin{eqnarray}    
&&\int{\bar W}(\{{\bf p}_i\})[n_1n_2(1-n_3)(1-n_4)-n_3n_4(1-n_1)(1-n_2)]\delta({\bf p}_1+{\bf p}_2-{\bf p}_3-{\bf p}_4-{\bf p}) \nonumber \\
&&\delta(\epsilon_1+\epsilon_2-\epsilon_3-\epsilon_4-\hbar\omega)\prod d{\bf p}_i.\label{ac1}
\end{eqnarray}
The quantity ${\bar W}$ is not necessarily the same as $W$. However, assuming that the angular integrations are the same at an arbitrary ratio $\hbar\omega/T$ and in the classical limit $T\gg\hbar\omega$, we may proceed with the integration over the energies. This gives:
\begin{eqnarray}    
&&\int [n_1n_2(1-n_3)(1-n_4)-n_3n_4(1-n_1)(1-n_2)]\delta({\bf p}_1+{\bf p}_2-{\bf p}_3-{\bf p}_4-{\bf p})\delta(\epsilon_1+\epsilon_2-\epsilon_3-\epsilon_4-\hbar\omega)\prod d\epsilon_i \nonumber \\
&&\propto T^2\omega\left[1+\frac{\omega^2}{4\pi^2T^2}\right].\label{ac2}
\end{eqnarray}
The absorption coefficient is proportional to this integral, and the proportionality coefficient (which depends only on $\omega$) can be found from the limiting case of $T\gg\hbar\omega$. So, the quantity in the square brackets in the rhs of Eq.(\ref{ac2}) represents the ratio of the damping rate of zero sound at an arbitrary value of $\hbar\omega/T$ to the damping rate at $T\gg\hbar\omega$. Using Eq.(\ref{tauT}) for $\tau_T$ we thus obtain the following damping rate at $T=0$:
\begin{equation}    \label{tau0}
\frac{1}{\tau_0}=\frac{\omega^2}{8\pi T^2}\frac{1}{\tau_{0T}}=\frac{2\epsilon_F}{\pi\hbar}\left(\frac{\hbar\omega}{\epsilon_F}\right)^2( p_Fr_*)^2D(\theta_0,\phi_k)=\frac{4\hbar}{m}n\left(\frac{\hbar\omega}{\epsilon_F}\right)^2(p_Fr_*)^2D(\theta_0,\phi_k), 
\end{equation}
and using Eq.(\ref{s0}) for $s_0$ we rewrite  $D(\theta_0,\phi_k)$ in the form:
$$D(\theta_0,\phi_k)=2(\cos^2\theta_0-\sin^2\theta_0\cos^2\phi_k)^2\ln
\left(\frac{1}{p_Fr_*}\right)+\tilde{\cal F}(\theta_0,\phi_k)-(\cos^2\theta_0-\sin^2\theta_0\cos^2\phi_k)^2
\ln\left(P_2^2(\cos\theta_0)+\frac{1}{8}\sin^2\theta_0\right).$$
\end{widetext}

The condition that the damping rate is much smaller than $(s_0-1)\omega$ requires the inequality
\begin{equation}  \label{c}
\frac{\hbar\omega}{\epsilon_F}\ll (p_Fr_*)^2,
\end{equation}
which is important for the visibility of the zero sound in the dynamical structure factor. 

The damping rate of zero sound is strongly anisotropic, and the anisotropy is similar to that of the relaxation rate of quasiparticles. The rate reaches maximum for dipoles perpendicular to the plane of their translational motion. We then have
$$D_{max}=2\ln\left(\frac{1}{p_Fr_*}\right),$$
and for $p_Fr_*\approx 0.5$ and $\epsilon_F/\hbar\omega\sim 100$ the damping time $\tau_0$ is by an order of magnitude larger than the relaxation time of quasiparticles with energy equal to $\hbar\omega$. The damping rate is minimal for dipoles lying in the plane of translational motion and the angle $\phi_k=\pi/2$. Then we obtain
$$D_{min}=\frac{4-\pi}{16}\simeq 0.05.$$   

\section{Concluding remarks}

The obtained results draw promising prospects for the observation of zero sound in 2D gases of polar molecules or magnetic atoms in the two-photon Bragg spectroscopy experiments by measuring the dynamical structure factor. This  becomes especially feasible in view of the recent success in creating spatially uniform ultracold quantum gases \cite{Hadzibabic}. The distance of the zero sound peak from the border of the particle-hole continuum (see Fig.2) is $\sim \omega (p_Fr_*)^4$. Comparing it with the damping rate of the zero sound given by equation (\ref{tau0}) we see that the latter is much smaller if the condition (\ref{c}) is satisfied. This condition is easily fulfilled even for rather small $p_Fr_*$. For realistic systems one can think of the zero sound frequency of the order of a few tens or hundreds of Hertz, whereas the Fermi energy can easily be a few kiloHertz (a few hundreds of nanokelvins), so that the ratio $\epsilon_F/\hbar\omega$ exceeds $10$. Under the condition (\ref{c}) the height of the zero sound peak in the structure function $(2\pi\hbar)^2S/m$ is $\sim \epsilon_F/\hbar\omega$, which is simply obtained replacing the $\delta$-function in Eq.(\ref{Ssnegative}) by $\tau_0 kv_F$. The maximum of the particle-hole continuum following from equation (\ref{Sspositive}) is $\sim (p_Fr_*)^{-2}$ and it is much lower under the condition (\ref{c}). Thus, the zero sound peak is not smeared out by the particle-hole continuum and can be visible in the dynamical structure factor. For example, if $p_Fr_*\approx 0.5$, then the separation between the border of the particle-hole continuum and the zero sound peak is $\sim\omega(p_Fr_*)^4\sim 20$ Hz for the sound frequency of a few hundred Hertz. It can be easily resolved as the relative frequency of the two Bragg beams can be controlled on the level of a Hertz.     

Owing to a remarkable progress in experiments with ultracold quantum gases, it is also promising to directly observe the propagation of zero sound in the 2D dipolar Fermi gas. Using a tightly-focused and far detuned laser beam one can create a potential to introduce a localized density modulation in the gas, without heating it. This technique has been used to directly study the propagation of sound in Bose-Einstein condensates \cite{FirstSound1} and in resonantly interacting Fermi gases \cite{FirstSound2,Grimm1,Grimm2}. In our system, the far detuned laser can be focused to the center of the 2D sample, and one can choose a proper power and shape of the excitation pulse to resonantly drive the desired zero sound mode (see, e.g. \cite{Grimm2}). After the zero sound mode is excited, one can observe the time evolution of the density profile and thus extract the information on the propagation of the mode.

Incoherent particle-hole excitations will also be excited during the pulse. However, as we have shown above, the decay of the zero sound is slower than that of particle-hole excitation.
After a time of the order of a fraction of the zero sound damping time $\tau_0$, let say $0.2\tau_0$ or $0.3\tau_0$, quasiparticle excitations are damped out and one is expected to see only the zero sound contribution to modulations of the density. The time $\tau_0$ can be easily made on the level of a second. For example,  this is the case for NaK molecules in the electric field providing $d\simeq 0.3$D ($r_*\simeq 50$nm). Then, at the 2D density $n\sim 10^9$ cm$^{-2}$ we have the Fermi energy approaching $1\mu$K and $p_Fr_*\approx  0.5$, and Eq.(\ref{tau0}) gives $\tau_0\sim 0.2$s for dipoles perpendicular to the plane of their translational motion.

\section*{Acknowledgements}

We are grateful to L.P. Pitaevskii for fruitful discussions. We acknowledge support from CNRS, from the IFRAF Institute, and from the Dutch Foundation FOM. This research was supported in part by the National Science Foundation under Grant No. NSF PHY11-25915. LPTMS is a mixed research unit No. 8626 of CNRS and Universit\'e Paris Sud.


\begin{thebibliography}{10}
\bibitem{shortnote} Strictly speaking, for $\phi_1$ close to $\phi+\pi$ we have to use the relation $\epsilon({\bf p}_i)=\hbar^2p_i^2/2m$ for the quasiparticle energy. Regarding all the rest, the calculations are similar to those for the case where $\phi_2$ is close to $\phi$ or to $\phi_1$.
\end{thebibliography}
\end{document}